\documentstyle[amsmath,amssymb,graphicx]{article}

\def\be{\begin{eqnarray}}
\def\ee{\end{eqnarray}}

\def\bfig{\begin{figure}[H] }
\def\efig{\end{figure}}

\def\bc{\begin{center}}
\def\ec{\end{center}}

\def\nn{\nonumber}

\def\p{\partial}
\def\lm{\limits}
\def\tr{{\rm tr}\,}
\def\Tr{{\rm Tr}\,}
\def\qtr{{\rm qtr}\,}

\hoffset= - 1.0in         

\title{{\bf On 3d extensions of AGT relation } \vspace{.2cm}}
\author{{\bf D.Galakhov}\thanks{{\small
{\it ITEP, Moscow, Russia and MIPT, Dolgoprudny, Russia}};
galakhov@itep.ru}, {\bf A.Mironov}\footnote{ {\small {\it Lebedev
Physics Institute} and {\it ITEP, Moscow, Russia}}; mironov@itep.ru;
mironov@lpi.ru}, {\bf A.Morozov}\thanks{{\small {\it ITEP, Moscow,
Russia}}; morozov@itep.ru}, {\bf A.Smirnov}\thanks{{\small {\it ITEP
Moscow, Russia and MIPT, Dolgoprudny, Russia}}; asmirnov@itep.ru}
\date{ }}

\begin{document}
 \maketitle

\vspace{-5.0cm}

\begin{center}
\hfill FIAN/TD-04/11\\
\hfill ITEP/TH-10/11\\
\end{center}

\vspace{3.5cm}

\centerline{ABSTRACT}

\bigskip

{\footnotesize
An extension of the AGT relation from two to three dimensions
begins from connecting the theory on domain wall
between some two S-dual SYM models
with the 3d Chern-Simons theory.
The simplest kind of such a relation would presumably connect
traces of the modular kernels in 2d conformal theory
with knot invariants.
Indeed, the both quantities are very similar,
especially if represented as integrals of
the products of quantum dilogarithm functions.
However, there are also various differences,
especially in the "conservation laws" for
integration variables, which hold for the monodromy traces,
but not for the knot invariants. We also discuss another possibility:
interpretation of knot invariants as solutions to
the Baxter equations for the relativistic Toda system.
This implies another AGT like relation: between
3d Chern-Simons theory and the Nekrasov-Shatashvili limit
of the 5d SYM.
}


\bigskip

\section{Introduction}

In \cite{TY} (see also \cite{3dAGTfirst}-\cite{3dAGTlast})
a long expected relation was suggested,
which can be considered as one of possible 3+3 counterparts
of the celebrated 2+4 AGT relation \cite{AGT} between
conformal blocks and Nekrasov functions.
This new relation is supposed to identify the
modular transformation kernels $M(a,a')$
of conformal blocks and matrix elements in
$3d$ Chern-Simons theory.
In its simplest version, relation is between the
trace of the modular kernel,
considered as a function of external dimensions,
and Hikami integrals \cite{HI},
representing the Chern-Simons partition functions
on $S^3/K$ (a $3d$ sphere with a knot $K$ removed)
considered as functions of monodromies around $K$.
A modest task of this letter
is to discuss possible formulations of such a
relation in an explicit form,
leaving aside all the general context and reasoning,
discussed in great detail in the profound text \cite{TY}.
We point out some problems with exact
identification of the modular trace
and the knot invariants.
We also attract attention to another
similarity: between the knot invariants and
$5d$ Seiberg-Witten theory,
which implies still another kind of
AGT relation, involving the
$3d$ Chern-Simons theory.

\section{Modular kernel}

The conformal block $B_\Gamma(a|m|q)$ for a given
graph $\Gamma$ depends on three kinds of variables:
$a$ and $m$ are parameters ($\alpha$-parameters)
on internal lines and external legs
respectively (the corresponding conformal dimensions
are quadratic in these parameters),
and $q$ parameterizes the graph itself.
The modular transformation does not change the graph $\Gamma$,
while changing $q\rightarrow q'$ and
permuting the entries within the set $\alpha$ of external-legs parameters
and is realized as an integral transform in $a$ variables:
\be
B_\Gamma(a|\alpha|q) = \int M(a,a') B_\Gamma(a'|\alpha'|q') d\mu(a')
\ee
The function $M(a,a')$ depends on $\Gamma$ and $\alpha$,
but not on $q$, and is called the modular kernel,
associated with the particular modular transformation
$q\rightarrow q'$.
In spirit, it is a Fourier kernel:
\be
M(a,a') \sim \exp \left(\frac{4\pi i aa'}{\epsilon_1\epsilon_2} + \ldots\right)
\ee
where the corrections are less singular at small $\epsilon_1$ and
$\epsilon_2$.

\bigskip

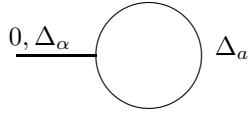
\begin{figure}\label{tor_bl_pc}
\begin{center}
\begin{picture}(75,50)
\put(0,25){\line(1,0){30}}
\put(50,25){\circle{60}}
\put(-3,29){$0, \Delta_\alpha$}
\put(75,25){$\Delta_a$}
\end{picture}
\end{center}
\caption{Toric conformal block diagram, $\Delta_a=\frac{\epsilon^2/4+a^2}
{\epsilon_1\epsilon_2}$\ , $\Delta_\alpha=\frac{\epsilon^2/4+\alpha^2}{\epsilon_1\epsilon_2}$\ .}
\end{figure}

The simplest example is provided by the modular transformation
of the toric 1-point function, Fig.1 which on the gauge theory side describes
the theory with adjoint matter with mass $m= -i\alpha + {\epsilon\over 2}
= -2i\tilde\alpha $:
\be
{\cal B}\big(a\big|\alpha\big|-1/\tau\big) =
\int {\cal M}(a,a'|\alpha){\cal B}\big(a'\big|\alpha\big|\tau\big)
d\mu(a')
\ee
Here \cite{Pog}
\be
{\cal B}(a|\alpha|\tau) = q^{-(\nu+1)/24}\eta(q)^{\nu}
e^{\frac{2\pi i\tau a^2}{\epsilon_1\epsilon_2}}
\left(1+
2q\frac{(\epsilon_1-m)(\epsilon_2-m)}{\epsilon_1\epsilon_2}
\frac{\big(\epsilon^2-4a^2+m(m-\epsilon)\big)}{(\epsilon^2-4a^2)} + O(q^2)\right)
\label{torblock}
\ee
with
$\nu = 1 - \frac{2m(\epsilon-m)}{\epsilon_1\epsilon_2}$,\
$\epsilon = \epsilon_1+\epsilon_2$ and
$\eta(q)=q^{\frac{1}{24}}\prod\lm_{k=1}^{\infty} (1-q^k)$, \ $q=e^{2\pi i\tau}$.
According to  \cite{Teschner} the modular kernel is given by
\be
{\cal M}(a,a'|\alpha) = \frac{2^{3/2}}{s(\alpha)}\int
\frac{s(a+r+\tilde \alpha)s(a-r+\tilde \alpha)}
{s(a+r-\tilde \alpha)s(a-r-\tilde \alpha)}\
e^{\frac{4\pi i ra'}{\epsilon_1\epsilon_2}}dr
\label{Mtor}
\ee
where the measure is
\be
d\mu(a')=4\sinh(2\pi\epsilon_1 a')\sinh(2\pi\epsilon_2 a')da'
\ee
The function $s(z)$ is the "quantum dilogarithm" \cite{diga,qdl},
the ratio of two digamma-functions \cite{diga},
\be
s(z|\epsilon_1,\epsilon_2) \sim
\prod_{m,n\geq 0}
\frac{\left(m+\frac{1}{2}\right)\epsilon_1
+\left(n+\frac{1}{2}\right)\epsilon_2 - i z}
{\left(m+\frac{1}{2}\right)\epsilon_1
+\left(n+\frac{1}{2}\right)\epsilon_2+i z}
\ee
Note that ${\cal M}(a,a'|\alpha)$ depends on the external momentum
$\tilde\alpha$,
the internal dimension
$\Delta(\alpha)$
and the central charge $c=1+6\frac{\epsilon^2}{\epsilon_1\epsilon_2}$,
but not on the modular parameter $\tau$.

\bigskip

When $\alpha\rightarrow 0$, the conformal block becomes pure classical:
${\cal B}(a|\alpha|\tau) \rightarrow
\frac{1}{\eta(q)}\ e^{2 i\pi\tau a^2/\epsilon_1\epsilon_2}$
(the last factor in the brackets in (\ref{torblock}) turns into $q^{1/12}\eta^{-2}(q)$ when $m=0$),
and the modular transformation turns into the ordinary Fourier transform:
\be
{\cal M}(a,a'|0)\mu'(a) \ \ \
\stackrel{\tilde\alpha=i\epsilon/2}{\Longrightarrow} \ \ \
\sqrt{2}\cos \left(4\pi i \frac{a a'}{\epsilon_1\epsilon_2}\right)\label{lim_mod}
\label{limM}
\ee
so that indeed
\be
\frac{e^{-2\pi i\tau^{-1} a^2/(\epsilon_1\epsilon_2)}}{\eta(-\tau^{-1})}=\int \frac{da'}{\sqrt{\epsilon_1\epsilon_2}} e^{4\pi i aa'/(\epsilon_1\epsilon_2)}\frac{e^{2\pi i\tau a'^2/(\epsilon_1\epsilon_2)}}{\eta(\tau)}
\ee
It is a somewhat non-trivial exercise to deduce (\ref{limM}) from eq.(\ref{Mtor}),
see Appendix A for the simplified case $\epsilon_1=-\epsilon_2=g_s$.

\bigskip

Notice that the monodromy kernel satisfies the unitarity relation
\be
\int d\mu(a) M(a,b) M^*(a,b')=\frac{d\mu(b')}{db'} \delta(b-b')
\ee
Therefore, it is natural to define the trace as
${\rm Tr}\sim \int \frac{d\mu(a)}{d\mu(a')}da'\delta(a-a')$.
Thus defined trace of the monodromy kernel (\ref{Mtor})
contains two integrals, but they actually split:
\be
\int {\cal M}(a,a|\alpha)da
= \frac{2^{3/2}}{s(\alpha)}\int\int
\frac{s(a+r+\tilde\alpha)s(a-r+\tilde\alpha)}
{s(a+r-\tilde\alpha)s(a-r-\tilde\alpha)}\
e^{4\pi i ra}dr da
= \frac{2^{3/2}}{s(\alpha)} T_+(\tilde\alpha)T_-(\tilde\alpha)
\ee
where
\be
T_\pm (\tilde\alpha) = \int\frac{s(z+\tilde\alpha)}{s(z-\tilde\alpha)}\
e^{\pm i\pi z^2}dz
\label{Tint}
\ee

In fact, the quantities like $T_\pm(\tilde\alpha)$
are well known from Chern-Simons theory,
and this opens a way towards $3d$ extensions of the
AGT conjecture.

\section{Exempts from the knot theory}
The polynomial knot invariants can be defined as averages of the Wilson loop along
the knot in the topological Chern-Simons (CS) theory \cite{CS}:
\be
\label{KA}
<K>_R = \left< \Tr\!_R P \!\exp\oint_K A\right>_{CS}
\ee
This invariant depends on the knot $K$, Lie algebra $G$, its representation $R$,
the CS coupling constant $\hbar = \log q = \frac{2\pi i}{k}$
(sometimes $k$ is shifted to $k+C_A$, as in the WZNW model \cite{WZNW}),
and, additionally, on the monodromy $u$, which describes
deviation from the periodicity of the field $A$ while circling around the knot.
One may think of $u$ as of the eigenvalue of $PSL(2)$ monodromy matrix around the knot $K$.
On the other hand, one may think that $u$ takes value
in the Cartan subalgebra of the gauge group $SU(2)$,
and, in this way, describes the representation $R$ living on the knot $K$.

The averages $<K>_R$ are clever generalizations of the ordinary
characters and, as all exact correlators, possess hidden
integrability properties \cite{UFN23,DV}. As a manifestation of this
hidden structure, the averages $<K>_R$ satisfy $K$-dependent
difference equations in the $R$(!) variable \cite{difeq}, which
allows one to consider them as belonging to the family of generalized
$q$-hypergeometric series. A $q\to 1$ limit of these equations defines
the spectral curve $\Sigma (K)$ and the saddle point of the corresponding
integral representation defines the associated Seiberg-Witten (SW) differential.
After that, the full $\hbar$-dependence can be
reconstructed with the help of the AMM/EO topological recursion
\cite{AMM/EO} from this SW data \cite{DFM}.

A nice property of invariants (\ref{KA}) is that at $u=0$ they are finite polynomials in $q=\exp(\frac{2\pi i}{k})$. In the literature, these polynomials normalized by the quantum dimension  have different conventional names corresponding to the choice of the group and representation. Here, for convenience we represent a table explaining the correspondence:
$$
\begin{array}{|c|c|c|}
\hline
&& \\
\textrm{Group}\backslash \textrm{Representation} & \textrm{Fundamental representation} & \textrm{General representation with weight} \ \ \lambda \\
&& \\
\hline
&& \\
N=0 &  \textrm{Convey-Alexander polynomial} & \textrm{--}\\ &&
\\\hline
&& \\
 SU(2) & \textrm{Jones polynomial} & \textrm{Colored Jones polynomial} \\
 && \\ \hline
&&\\
SU(N) & \textrm{HOMFLY polynomial} & \textrm{Colored HOMFLY polynomial}\\
&& \\
\hline
&& \\
SO(N) &  \textrm{Kauffmann polynomial} & \textrm{Colored Kauffman polynomial}\\ &&
\\
\hline
&& \\
\left\{SU(N)\right\}_t &  \textrm{superpolynomial} & \textrm{Colored superpolynomial}\\ &&
\\
\hline
\end{array}
$$
The last line of this table describes an extension from (quantum) groups to MacDonald
characters which leads to a one-parametric deformation ($t$-deformation) of (\ref{KA}),
to superpolynomials \cite{sp,GIKV} involving the Khovanov homology
\cite{Hov}. Further extension from MacDonald to the Askey-Wilson-Kerov level remains
untouched so far.

\bigskip

Six kinds of representations are currently known
for the Wilson averages in Chern-Simons theory,
we briefly describe them in the following
subsections and provide some explicit examples
in Appendix B.

\subsection{Representation through quantum $R$-matrix \cite{QRM}}
This representation appears when calculations in CS theory are done in
the temporal gauge $A_0=0$ \cite{MoSm}.
Then the propagator is ultralocal, and only the crossings ($c$)
and extremal ($e$) points
in projection of $K$ on the $xy$ plane contribute,
and the answer can be schematically written as
\be
\label{tf}
<K>_R = \Tr\!_R \overrightarrow{\prod_{e,c}}\
{\cal U}_e {\cal R}_c
\ee
where the ordered product is taken along the line $K$,
${\cal{R}}$ is the quantum $R$-matrix in representation $R$,
and ${\cal{U}} = q^\rho$ is the "enhancement" of the ${\cal R}$-matrix
in the same representation.
For the braid representation of the knot,
this formula reduces to the well-known formula for the
quantum group invariants of knot:
\be
\label{qi}
<K>_R =q^{-w(
b_{K})\,\Omega_{2} (R) }\,\qtr_{R}( b_{K} )
\ee
Here $b_{K}\in B_{n}$ is the element of the braid group,
representing the knot $K$, i.e. its closure gives $K$,
$\Omega_{2}(R)=\tr_{R}(T^a T^{a})/\dim R$ is the value of quadratic Casimir
function in the representation $R$
and $\qtr_{R}$ is the quantum trace over the representation $R$,
$\qtr_R b_K = \tr_R b^K q^{\rho^{\otimes n}}$.
The function $w(b_{k})$ is the so called writhe number of the braid $b_{K}$,
it is equal to the total sum of orientations of the crossings:
$$
b_{K}=\prod\limits_{\{k\}} g_{k}^{n_k} \ \ \ \Longrightarrow\ \ \
w(b_K)=\sum\limits_{\{k\}} \, n_{k}
$$

Calculation of (\ref{qi}) for some particular knot is just a matter of
multiplication and taking a trace of relatively big matrices.
The braid representations of several first knots and the writhe
numbers of the corresponding closures are summarized in the following table:
\be \label{kntable}
\begin{array}{|l|l|l|}\hline \textrm{Knot}& \textrm{Braid representation} &
\textrm{Writhe} \\ \hline 3_1 & b=g_{1}^{3}\in B_{2} & w({\hat b})=3 \\
\hline 4_{1} & b=g_{2}2\, g_{1}^{-1}\, g_{2} \, g_{1}^{-1}\in
B_{3} & w( {\hat b} )=1\\ \hline 5_{1} & b=g_{1}^{5}\in B_{2} &
w({\hat b})=5\\ \hline 5_{2} &
b=g_{2}^{3}\,g_{1}\,g_{2}^{-1}\,g_{1} \in B_{3} & w(\hat b)=4\\
\hline 6_{1} & b=
g_{1}\,g_{2}^{-1}\,g_{3}\,g_{1}\,g_{2}^{-1}\,g_{3}^{-2}\in B_{4} &
w({\hat b})=-1\\ \hline
\end{array} \ee\\

\subsection{Representation through classical $R$-matrix
and quantum associator \cite{QA}}

A similar representation for $<K>_R$ trough classical instead of quantum $R$-matrices
appears in the calculation of (\ref{KA}) in the holomorphic gauge $A_{\bar{z}}=0$.
However, instead of the trivial insertions of $q^\rho$ factors one now needs
to insert sophisticated Drinfeld associators \cite{quas}.
In the holomorphic gauge the theory reduces to the Kontsevich integral of knot \cite{Konint}.
In this case, one can construct the representation of the braid group $B_{n}$
in $R^{\otimes n}$ though the Drinfeld associators as follows: for the element
$g_{k}\in B_{n}$ one has:
\be
g_{k}\rightarrow \Psi_{k} R_{k} \, \Psi_{k}^{-1},\ \ R_{i}= \underbrace{1\otimes1
\otimes...\otimes1}_{1...i-1}\,\otimes\,R
\otimes \underbrace{ 1 \otimes...\otimes1}_{i+2...n}, \ \  \textrm{with} \ \  R=q^{T^{a} \otimes T^{a}  }
\ee
$\Psi_{k}=\Phi_{k}\otimes 1^{\otimes (n- k)}$, where $\Phi_{k}$ is the $k$-th Drinfeld associator.
Then, again, if $b_{K}\in B_{n}$ is the braid representing some knot $K$,
for the quantum invariant (\ref{KA}) one obtains
\be
\label{KI}
<K>_R = \qtr_{R}(1)^{n}\,q^{-w(
b_{K})\,\Omega_{2} (R) }\,\qtr_{R}( b_{K} )
\ee
Again, the classical $R$-matrices $R=q^{T^{a} \otimes T^{a}}$  stand just at the intersection
points of the $K$ projection on a plane.
However, instead of the simple enhancement factors $q^\rho$,
one now inserts in between the quantum associators \cite{quas},
acting on $k$ lines simultaneously.
The Drinfeld associators are solutions
to the Knizhnik-Zamolodchikov equations \cite{KZ} in the
WZNW conformal theory \cite{WZNW}.
In particular, the $k$-th associator describes the monodromy
of $k+2$-point function in the WZNW model, and the Kontsevich integral (\ref{KI})
is nothing but the trace of monodromy associated with the braid $b_{K}$.
For further details of this representation see
\cite{DBetal}.

\subsection{Representation through Vassiliev invariants
and Kontsevich integrals}

In CS theory perturbative expansion, the dependencies of the Wilson average
$<K>_R$ on the knot $K$ and on the group structure $G,R$ are nicely separated:
\be \label{vr}
<K>_{R}\,=\dim_q(R)\prod\limits_{m=0}^{\infty}\,\prod_{n=1}^{d_{m}}\,
\exp\Big( \hbar^m \alpha_{m,n}(K)\,r_{m,n}(R) \Big)
\ee
where $dim_{q}(R)$ is the quantum dimension of representation:
$$
\dim_{q}(R)=\dfrac{q^{{N\over 2}}-q^{-{N\over 2}}}{q^{1\over 2}-q^{-{1\over 2}}}
$$
The quantities $\alpha_{m,n}(K)$ are the primary Vassiliev invariants \cite{vass},
which are rational(!) numbers,
naturally represented either as modifications of the Gauss linking integrals in the Lorentz
gauge $\partial_\mu A_\mu=0$ \cite{Bar}, or as the Kontsevich integrals \cite{Konint} in
the holomorphic gauge $A_{\bar z}=0$, or through the writhe numbers \cite{TurVir} in the
temporal gauge $A_0=0$.

\begin{figure}
\begin{center}
\includegraphics[scale=0.6]{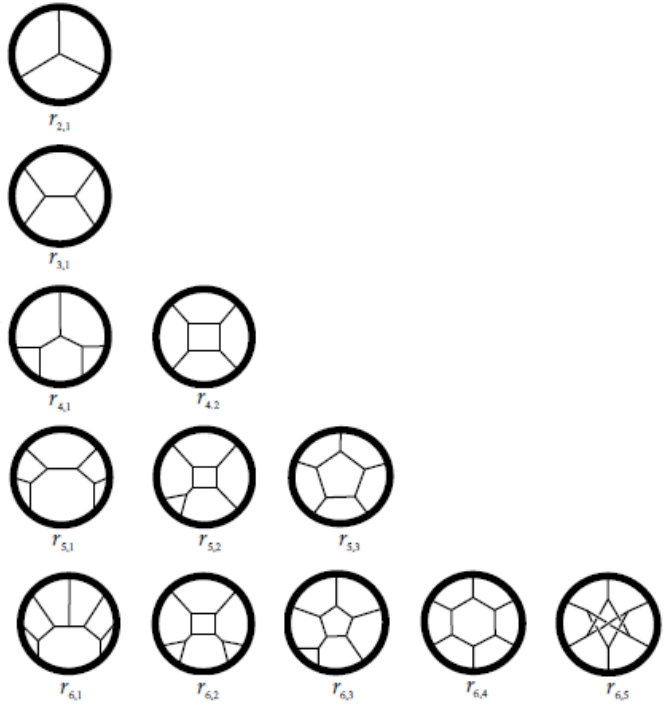}
\caption{A basis of independent Casimir function up to degree $6$.\label{cas}}
\end{center}
\end{figure}

The group factors $r_{m,n}$ are the
eigenvalues of operators
from the cut-and-join algebra \cite{MMN} on the $GL(\infty)$ characters $\chi_R$,
and form a basis of the
\textit{multiplicative independent} Casimir eigenvalues of order $m$.
In every order $m$ one has $d_{m}$ independent Casimirs, the first several values are:
\begin{equation}
\begin{array}{|c|c|c|c|c|c|c|}
\hline
m&1&2&3&4&5&6\\
\hline
d_{m}&0&1&1&2&3&5\\
\hline
\end{array}
\end{equation}
\noindent
$r_{kj}(R)$ can be also understood as eigenvalues of operators
from the cut-and-join algebra \cite{MMN} on the characters $\chi_R$,
and are certain non-linear combinations of the Casimir eigenvalues.
The Casimir functions $r_{m,n}$ are polynomials of the rank and the weight
of the representations, the following table lists the first four Casimirs for $SU(n)$, $SO(n)$
in the fundamental representation and for $SU(2)$ in the spin $J$ representation:
\be
\label{cfun}
\begin{array}{|l|l|l|l|l|}
\hline
 & r_{2,1} & r_{3,1} & r_{4,1} & r_{4,2}\\
 \hline &&&& \\
SU(n)\ \ \textrm{fund.}& -\frac{n^2-1}{4}&
-\frac{n (n^2-1)}{8}&-\frac{n^2 (n^2-1)}{16} & -\frac{(n^2+2) (n^2-1)}{16}\\
&&&& \\
SO(n)\ \ \textrm{fund.}& -\frac{(n-2) (n-1)}{16}  &-\frac{(n-2)^2 (n-1)}{64}&-\frac{(n-2)^3
(n-1)}{256}& \frac{\left( n-1 \right)  \left( n-2 \right)  \left( {n}^{2}-5\,n+10 \right)}{256}\\
&&&& \\
SU(2),\ \ J &-J(J+1)&-J(J+1)&-J(J+1)& 2 J^2 (J+1)^2  \\
&&&& \\
\hline
\end{array}
\ee
Our choice of the basis of $r_{m,n}$ is shown in Fig.\ref{cas}
in the form of the chord diagrams:
the circle stands for the trace over representation $R$,
and the trivalent vertices depict the structure constants $f^{a b c}$ of the algebra.
This notation is clear from the two examples:
$$
r_{2,1}=\frac{1}{\dim_R} f^{a b c}\tr_{R}(T^{a} T^{b} T^{c}),\ \ \
\ \ r_{3,1}= \dfrac{1}{\dim_R }f^{a b e} f^{e c,d}\tr_{R}(T^{a} T^{b} T^{d} T^{c})
$$

In the case of $G=SL(2)$ the relevant representations $R$
are labeled by the value of spin $J$, and $r_{kj}(J)$ are
polynomials of $J$ of degree $k$.
The series in the exponential (\ref{vr}) can be re-expanded
in new variables $\hbar$ and $N = 2J+1=u/\hbar$:
\be
F(K,\hbar,u) =\log <K>_{R}-\log N= \sum_{k=0}^\infty \hbar ^k \sum_{j=1}^{d_k}
\alpha_{kj}(K) r_{kj}(N/\hbar) =
\sum_k \hbar^k F_{k}(K,u)
\label{Fdef}
\ee
These $F_k(K,u)$ are infinite combinations of Vassiliev invariants of all orders.
The $K-J\ $ separation gets obscure after such a re-expansion,
and all the information about the gauge group ($r_{m,n}$) and the knot ($\alpha_{m,n}$)
gets mixed.
Instead this expansion is the knot theory counterpart of the
genus ($1/N$) expansion in matrix models, and $u = \hbar N$ plays role
of the t'Hooft's coupling.

Remarkably, for some knots (called {\it hyperbolic})
$F(K)$ behaves as $V(K,u)\hbar^{-1}$
at small $\hbar$ \cite{Kash}, this is despite all terms in the series
have positive powers of $\hbar$!
This is typical for genus expansion series:
contributions of every particular genus are all convergent series,
but the sum over genera diverges:
it can be studied with the help of the Pade summation methods,
see \cite{Marino,MoShaHZ} for a recent description within the
matrix model context.
The value of the coefficient $V(K,u=1)$ at $u=1$ coincides
with the volume of $S^3/K$ in the uniquely defined hyperbolic metric \cite{Turst},
the situation is far more interesting for non-hyperbolic knots
({\it toric} or {\it satellite}),
see the plots in appendix C.

\subsection{Representation through quantum dilogarithm \label{RepQL}}

In \cite{difeq} the polynomial invariants (\ref{KA})
for the gauge  group $SU(2)$ with spin $J$
were systematically interpreted as generalized $q$-hypergeometric functions,
such representations were widely used before in particular examples.
This means that these Wilson averages can be represented in the form
of finite sums:
\be
\label{pr}
<K>_{J}=\dim_{q}(R)\sum\limits_{k_{1},k_{2},...,k_{n}}
\dfrac{(q,N)_{k_1}...(q,N)_{i_{1}}}{(q,N)_{k_{i_2}}...(q,N)_{k_n}} \,q^{p_{2}(k_1,...,k_{n})}
\ee
Here $N=2J+1$,  $(q,N)_{k}$ is the $q$-Pochhammer symbol:
\be
(q,N)_{k}=\prod\limits_{i=1}^{k} \Big( q^{\frac{N-i}{2}}- q^{-\frac{N-i}{2}} \Big)\
\ee
and $p_{2}(k_1,...,k_{n})$ is a certain quadratic function.
The numbers of $k_{i}$ in the sum (\ref{pr}), the quadratic polynomial $p_{2}$,
the arrangement of the Pochhammer symbols are determined by the knot $K$
and the limits of summation by $J$ .
The existence of such a representation for knots directly follows
from the $AJ$-conjecture for the colored Jones polynomials \cite{AJconj},
which states that $<K>_{j}$ is a solution of certain hypergeometric difference equation,
see also s.\ref{SC and TR} below.
At the same time, the Pochhammer symbols can be expressed through the ratio of
the quantum dilogarithms:
\be
(q,N)_{k}=(-i)^k\, \dfrac{s\left( i \epsilon_2 (N-1-k)
+ \frac{i}{2} (\epsilon_1+\epsilon_2) \Big| \epsilon_1, \epsilon_2 \right)}
{s\left(i \epsilon_2 (N-1) + \frac{i}{2} (\epsilon_1+\epsilon_2) \Big| \epsilon_1, \epsilon_2 \right)}
\ee
and the expression (\ref{pr}) can be schematically rewritten as
\be
\label{pr1}
<K>_{J}=\dim_q(R)\sum\limits_{k_{1},k_{2},...,k_{n}} \dfrac{s_{k_1}...s_{k_{i_1}} }{s_{k_{i_2}}...s_{k_n}}
\,q^{p_{2}(k_1,...,k_{n})}
\ee
This provides a discrete version of the Hikami invariants.

\subsection{Hikami formalism in Chern-Simons theory}
The idea of Hikami formalism is to calculate the CS partition functions,
making use of triangulation of the $3d$ manifold $M$, i.e. by
decomposing it into elementary symplices, tetrahedra.
Each tetrahedron has four faces, one can choose two and call them
white, then the other two will be black, and one glues
black faces to the white ones only.
With each site one associates a number $p$ with the white faces,
and $p^*$ with the black ones. When two sites are identified,
the two numbers are identified, $p_i = p_j^*$.
Finally, with each tetrahedron one associates a function
$G(p_1^*,p_2^*|p_1,p_2)$, and integrates over all $p$-variables
on identified faces.
In this way, with each triangulation of $S^3$ one associates
a multiple integral over all $p$-variables, one per each $2$-face
of triangulation:
\be
H(K|u) = \int \prod_{simplices} dp_{i_1}dp_{i_2} dp_{i_1}dp_{i_2}
G(p_{i_1}^*,p_{i_2}^*|p_{i_1},p_{i_2})
\prod_{2-faces} \delta_{p^*_{k_m}-p_{k_n}}
\prod_{1-cycles} \delta\left(\sum p_j - u\right)
\label{BGint1}
\ee
At the same time, geometrically, one can associate with the
system of the glued tetrahedron a knot $K$, moreover,
the $u$-variables, which are associated with the 1-cycles,
can be interpreted as the $U(1)$-monodromies around the knot $K$.
Most important, the functions $G$ were found in \cite{HI} to be
\be
G(p_{i_1}^*,p_{i_2}^*|p_{i_1},p_{i_2}) =
\delta(p_1+p_2-p^*_1) s\big(p_2^*-p_2 - i\pi + \hbar\big|
\epsilon_1,\epsilon_2\big)
e^{p_1(p_2^*-p_2)/2\hbar + \hbar^2/6}
\ee
and the Hikami integral $H(K|u)$ for the given $K$ has exactly the same form as
the Wilson average $<K>_R$ from s.\ref{RepQL}
in the $(\hbar, u)$ variables,
only the sum is substituted by the integral, see examples in Appendix B below.
The transition from sums to integrals is associated with the switching from
compact to non-compact groups.

\subsection{Spectral curves and topological recursion\label{SC and TR}}

At least for $G=SL(2)$
the average $<K>_R$ is annihilated by a $K$-dependent difference operator,
i.e. satisfies a recurrent relation in spin $J$, this statement
is sometime called the AJ-conjecture \cite{AJconj} (for the first examples beyond
$SL(2)$ see \cite{AJ3}).
In the variables $u=N\hbar$, where $N=D_{2J+1}=2J+1$
one has:
\be
{\cal A}(e^{\hbar\p_u}, e^u) <K>_R = 0
\label{AJeq}
\ee
In the $\hbar=0$ limit this operator turns into a function (polynomial),
and the difference equation into an algebraic one,
\be
\Sigma(K):\ \ \
{\cal A}(w,\lambda) = 0
\ee
defining the spectral curve $\Sigma(K)$. One can further define the Seiberg-Witten differential
\be
dS=\log w d\log\lambda
\ee
In the typical examples of $4_1$ and $m009$ knots, this is a quadratic equation
in $w$ (note that an additional $U(1)$ factor $w-1$ splits away decreasing degree
of the equations by one):
\be
\Sigma(K):\ \ \
{\cal A}(w,\lambda) = w+\frac{1}{w}
= 2f\left(\lambda+{1\over\lambda}\right), \ \ \ \
w_\pm(\lambda) = f\pm \sqrt{f^2-1}
\label{spec}
\ee
and
\be
dS = \Big(\log w_+(\lambda) - \log w_-(\lambda)\Big)d\log \lambda
= \log\frac{f+\sqrt{f^2-1}}
{f-\sqrt{f^2-1}}\ d\log\lambda
\label{diff}
\ee
The free energy (\ref{Fdef}) $F = \log<K>_R$
is actually an exact Seiberg-Witten prepotential,
reconstructed in all orders of genus expansion in $\hbar$
from this data $\big(\Sigma_K, dS\big)$
with the help of the AMM/EO topological recursion \cite{AMM/EO}.
See \cite{DFM} for detailed examples of this reconstruction.
One may wonder, {\it what} is the corresponding SW theory, --
and one easily recognizes in (\ref{spec}) and (\ref{diff})
the formulas from \cite{5d}, describing the $5d$ version
of the classical SW theory in terms of the relativistic Toda integrable system.
Then (\ref{AJeq}) should be {\it a} Baxter equation,
which is now known \cite{MMns,NSmore}
to describe the NS deformation of the classical SW theory
(or, what is the same, the NS limit of the full SW theory).
This observation implies a new kind of AGT duality,
to be discussed in s.5 below.
When ${\cal A}(w,\lambda)=0$ is not reduced to a quadratic
equation in $w$,
the analysis is more complicated.

\section{A $3d$ AGT relation}

Suggestion of \cite{TY} is somehow to identify
the modular kernels $M(a,a')$, associated with
modular transformations of the punctured Riemann
surface $S(q)\longrightarrow S(q')$,
with Chern-Simons cobordism amplitudes on the $3d$ space,
which interpolates between $S(q)$ and $S(q')$.
In particular, the trace of $M(a,a')$ should be compared
with the knot invariants,
where the relevant knot $K$ is formed by the closed
trajectories of punctures, while external momenta are associated
with monodromies around the knot components.
Of course, $K$ depends on the choice of the modular
transformation.
Indeed, there is a striking similarity between
both types of quantities, (\ref{1}) and (\ref{2}) below.
Both are multiple integrals of products of the
quantum-dilogarithm functions $s(\ldots|\epsilon_1,\epsilon_2)$,
and there is a natural identification of parameters:
\be
2\pi i\hbar = \log q = \frac{2\pi i}{k+C_G}\ \ {\stackrel{\cite{TY}}{=}}\ \
2 \pi i b^2 = \frac{2\pi i\epsilon_2}{\epsilon_1}
\label{AGTcorr}
\ee
However, there are also differences:
in the number of $s$-functions,
in their arguments and in the integration contours.
Perhaps, the most striking difference is that the
integration variables obey "conservation laws" in
expressions for know invariants,
but they do not do so in the trace of monodromy matrix.
To see this, one can use the property
\be
s(z) = 1/s(-z)
\ee
to put all the $s$-functions in the numerator,
so that all the integrals acquire the "canonical" form:
\be
\prod_i \int_{C_i} dp_i \prod_m s(A_{mi}p_i+B_m|\epsilon_1,\epsilon_2)
\exp\left(C_{ij}p_ip_j + D_ip_i +E\right)
\ee
Then
\be
\sum_m A_{mi} = 0
\ee
in (\ref{Tint}),
but this is not the case  for knot invariants, at least for
some $i$, i.e. for some of the integration variables.
The simplest example of such a discrepancy is between
\be\label{1}
T_\pm(z) = \int \frac{s(z+\tilde\alpha)}{s(z-\tilde\alpha)}e^{\pm i\pi z^2}
\ee
which are constituents in the expression for the toric modular trace, and
\be\label{2}
<4>_1 \sim \int s(z+u)s(z-u) e^{\frac{6\pi i uz }{\epsilon_1\epsilon_2}} dz
\ee
for the Wilson/Hikami average, associated with the $3d$AGT-related $4_1$ knot.
Thus, explicit formulation of the $3d$ AGT hypothesis
is clearly very near, but still remains to be found.

\section{A route to alternative AGT relation}

A seeming failure of the $3+3$ AGT relation
can attract some new attention to possible alternatives:
one actually expects a lot of different AGT like
relations to exist.
An obvious possibility is to look for an extension
of $2+4$ to $3+5$, and try to relate quantities in
the $3d$ Chern-Simons theory with those in
the $5d$ SYM.
This is of course a far more straightforward
exercise.
Equations in section \ref{SC and TR} above describe
knot invariants, but it is easy to recognize in
(\ref{spec}) the spectral curve for the relativistic Toda system,
and in (\ref{diff}) the corresponding Seiberg-Witten
differential \cite{5d}.
Thus, these equations establish a clear link between
$3d$ Chern-Simons and the $5d$ version of Seiberg-Witten
theory described as a straightforward $q$-deformation
of the ordinary $4d$ SW theory.

Within this identification, one should associate the corresponding
difference equation (\ref{AJeq}) with the Baxter equation for the same system.
The Baxter equations are known \cite{MMns}
to arise in the Nekrasov-Shatashvili
limit \cite{NS} and, therefore, one obtains a new AGT like
relation:

\be\boxed{
3d \ {\rm Chern-Simons}
\stackrel{new\ AGT}{\longleftrightarrow}
{\rm NS\ limit\ of}\ 5d \ {\rm LMNS\ prepotential}
\stackrel{ordinary\ AGT\ \cite{5dMMSm}}{\longleftrightarrow}
q-{\rm Virasoro\ conformal\ blocks}}
\nn
\ee

\bigskip

More concretely, in accordance with \cite{MMns}, one associates with
the solution to the Baxter equation, i.e. with $<K>_R$, the SW
differential, its monodromies around the $A$- and $B$-cycles on the
spectral curve (\ref{spec}) giving rise to the $5d$ Nekrasov
functions in the NS limit via the SW equations.

There are a few interesting points to be mentioned already
at this stage.
The dilogarithm formulae for the knot invariants from ss.3.4 and 3.5
provide integral representations for solutions of
equations (\ref{AJeq}).
They look very similar to solutions \cite{Khar}
for the {\it open} quantum relativistic Toda chain system,
and such solutions are {\it not} available
for the {\it closed} chain, but (\ref{spec})
is of this latter kind!
The thing is that, first, (\ref{AJeq})
defines a Baxter equation at some special point of the
moduli space: for special values of "energies",
thus, the fact that solutions are unavailable in such
a form at a generic point does not forbid them to exist
at some special point.
Second, while the classical equations (\ref{spec})
are clearly of the relativistic Toda type,
their quantization is not unique, and (\ref{AJeq})
is not the standard version of the Baxter equation,
considered in the literature.
One more important point is that when dealing with
difference (rather than differential) equations,
one obtains infinitely more solutions.
To fix this ambiguity one imposes two difference equations
where one would impose only one differential.
And this pair of equations is typically related by the
symmetry $\epsilon_1\leftrightarrow \epsilon_2$,
which is explicitly broken by all our construction
of knot invariants. In particular, it is explicitly
broken in the basic AGT identification (\ref{AGTcorr}).

An interesting question is what should substitute the
knot invariants when this new AGT duality is lifted
to the entire LMNS deformation of $5d$ theory, beyond the
Nekrasov-Shatashvili limit.

\section{Conclusion}

After discovery of the AGT relation \cite{AGT},
which embeds the $2d$ conformal theory
into the general context of Seiberg-Witten
and integrability theory \cite{SWint},
a hunt has immediately started
for its extension, which would do the
same with the $3d$ Chern-Simons theory.
The goal of this letter
is to switch the discussion of the $3d$ AGT
relations from the qualitative to a quantitative
mode.
This is made possible by the extensive progress
in the theory of knot invariants,
which is briefly reviewed in s.3
of the present letter.
(Actually, some derivations from CS theory
itself are still lacking, but this is mostly
because of the insufficient attention
to these important problems within the
QFT community.)
Given the existing set of explicit formulas,
one can easily test various suggestions.
In this way we point out some problems
with the suggestion by \cite{TY} to
AGT-relate the knot invariants with
the modular kernels.
Instead, we demonstrate that these invariants
are AGT-related to the $5d$ SYM theory,
this is a more straightforward and less
intriguing option, still it also deserves
the attention.

\section*{Acknowledgements}
We acknowledge useful consultations from P.Dunin-Barkovsky,
S.Kharchev and A.Sleptsov. Our work is partly supported by Ministry
of Education and Science of the Russian Federation under contract
02.740.11.5194, by RFBR grants 10-01-00536 and by joint grants
11-02-90453-Ukr, 09-02-93105-CNRSL, 09-02-91005-ANF,
10-02-92109-Yaf-a, 11-01-92612-Royal Society.

\newpage

\section*{Appendix A.  Dilogarithm properties}

A big problem with discussions of $3d$ AGT relations
is the lack of common notations:
people coming from different fields
use definitions of the same quantities,
which differ by all kinds of rescalings.
The purpose of this Appendix is to list at
least some relations between various definitions of
quantum dilogarithms used in the literature: $S(z|\epsilon_1,\epsilon_2)$,
$S_b(z)$, $\Phi_h(z)$, $e_b(z)$.
We also demonstrate the trick needed
to take the massless limit $\alpha\rightarrow 0$
of the modular kernel $M(a,a')$,
i.e. to derive eq.(\ref{lim_mod}).

\subsection*{A1. Various dilogarithms}

The function $s(z)$ is the "quantum dilogarithm" \cite{qdl},
the ratio of two digamma-functions \cite{diga},
\be
s(z|\epsilon_1,\epsilon_2)=
\prod_{m,n\geq 0}
\frac{\left(m+\frac{1}{2}\right)\epsilon_1
+\left(n+\frac{1}{2}\right)\epsilon_2 - i z}
{\left(m+\frac{1}{2}\right)\epsilon_1
+\left(n+\frac{1}{2}\right)\epsilon_2+i z}
= \frac{\Gamma_2(\epsilon/2+iz|\epsilon_1,\epsilon_2)}
{\Gamma_2(\epsilon/2-iz|\epsilon_1,\epsilon_2)}
\ee
It enjoys a set of periodicity properties
\be
s\left(z-\frac{i\epsilon_2}{2}\Big|\epsilon_1,\epsilon_2\right)=2\cosh\left(\frac{\pi z}{\epsilon_1}\right)s\left(z+\frac{i\epsilon_2}{2}\Big|\epsilon_1,\epsilon_2\right)\\
s\left(z-\frac{i\epsilon_1}{2}\Big|\epsilon_1,\epsilon_2\right)=2\cosh\left(\frac{\pi z}{\epsilon_2}\right)s\left(z+\frac{i\epsilon_1}{2}\Big|\epsilon_1,\epsilon_2\right)\\
s\left(z-\frac{i\epsilon}{2}\Big|\epsilon_1,\epsilon_2\right)=4\sinh\left(\frac{\pi z}{\epsilon_1}\right)\sinh\left(\frac{\pi z}{\epsilon_2}\right)s\left(z+\frac{i\epsilon}{2}\Big|\epsilon_1,\epsilon_2\right)
\label{shift_s}\label{s_rel}
\ee
This definition of the quantum dilogarithm admits the integral representation
\be
i\log s(z|\epsilon_1,\epsilon_2)
=\int\lm_0^{\infty}\frac{dw}{w}\left\{\frac{\sin (2zw)}
{2\sinh (\epsilon_1 w)\sinh (\epsilon_2 w)}-\frac{z}{\epsilon_1\epsilon_2 w}\right\}
\ee
which can be used to derive the asymptotic formulae
\be\label{expan}
i\log s(z|\epsilon_1,\epsilon_2)
=\frac{\pi z^2}{2\epsilon_1\epsilon_2}
-\frac{\pi}{24}\frac{2\epsilon_1\epsilon_2-\epsilon^2}{\epsilon_1\epsilon_2}
+i\sum\lm_{n=0}^{\infty}\frac{B_n(1/2)}{n!}
\left(2\pi i\frac{\epsilon_2}{\epsilon_1}\right)^{n-1}
{\rm Li}_{2-n}\left(-e^{\frac{2\pi z}{\epsilon_1}}\right)
\ee
The following resummation expansion is also of use
\be
i\log s(z_0+z|\epsilon_1,\epsilon_2)=\frac{\pi(z_0+z)^2}{2\epsilon_1\epsilon_2}-\frac{\pi}{24}\frac{2\epsilon_1\epsilon_2-\epsilon^2}{\epsilon_1\epsilon_2}
+\sum\lm_{k=-1}^{\infty}\sum\lm_{j=0}^{\infty}\frac{i^{k+1}B_{k+1}(1/2)}{(k+1)! j!}(2\pi)^{k+j}\frac{\epsilon_2^k}{\epsilon_1^{k+j}}{\rm Li}_{1-j-k}\left(-e^{\frac{2\pi z_0}{\epsilon_1}}\right)z^j
\ee
The quantum dilogarithm is symmetric w.r.t. $\epsilon_{1}$ and $\epsilon_2$, however,
we have explicitly chosen $\epsilon_2$ to be small here, this expansion playing a crucial role
in the NS limit.

This definition of the quantum dilogarithm is convenient for applications in the context of
AGT conjecture, where parameters $\epsilon_{1,2}$ are explicitly specified. Though in the
literature there is another definition which differs by rescaling
\be
S_b(z)=\exp\left(\frac{1}{i}\int\lm_{0}^{\infty}\frac{d w}{w}\left(\frac{\sin 2z w}{2\sinh bw \sinh b^{-1}w}-\frac{z}{w}\right)\right)
\ee
so that
\be
\boxed{s(z|\epsilon_1,\epsilon_2)=S_{\sqrt{\frac{\epsilon_1}{\epsilon_2}}}\left(\frac{z}{\sqrt{\epsilon_1\epsilon_2}}\right)}
\ee

\subsection*{A2. Modular kernel}

In this section we present a standard trick to compute the modular kernel (\ref{Mtor}) in the simple limit of zero external dimension $\tilde \alpha=i\frac{\epsilon}{2}$.
Considering the modular kernel in this limit naively, one derives from
eq.(\ref{s_rel}) that
\be
M=2^{3/2}\int dr \frac{e^{4\pi i a'r}}{16\sinh\left(\frac{\pi (a+r)}{\epsilon_1}\right)\sinh\left(\frac{\pi (a+r)}{\epsilon_2}\right)\sinh\left(\frac{\pi (a-r)}{\epsilon_1}\right)\sinh\left(\frac{\pi (a-r)}{\epsilon_2}\right)}
\ee
The denominator in the integrand has double poles, which were glued together in the limit
$\tilde\alpha\rightarrow i\frac{\epsilon}{2}$, and the chosen integration contour is pinched between
them. Hence, one has to take the limit more carefully:
\be
{\cal M}(a,a'|0)=\oint\lm_{r=-a} dr \frac{s(a+r+i\epsilon/2+i\lambda)}{s(a+r-i\epsilon/2-i\lambda)}\frac{e^{4\pi i a'r}}{4\sinh\left(\frac{\pi (a-r)}{\epsilon_1}\right)\sinh\left(\frac{\pi (a-r)}{\epsilon_2}\right)}+\nn\\
+\oint\lm_{r=a} dr \frac{s(a-r+i\epsilon/2+i\lambda)}{s(a-r-i\epsilon/2-i\lambda)}\frac{e^{4\pi i a'r}}{4\sinh\left(\frac{\pi (a+r)}{\epsilon_1}\right)\sinh\left(\frac{\pi (a+r)}{\epsilon_2}\right)}
\ee
Since
\be
\frac{s(a+r+i\epsilon/2+i\lambda)}{s(a+r-i\epsilon/2-i\lambda)}=\prod\lm_{m,n\geq 0}\frac{(m+1/2)\epsilon_1+(n+1/2)\epsilon_2-i(a+r+i\epsilon/2+i\lambda)}{(m+1/2)\epsilon_1+(n+1/2)\epsilon_2+i(a+r+i\epsilon/2+i\lambda)}
\nn\times\\ \times\frac{(m+1/2)\epsilon_1+(n+1/2)\epsilon_2+i(a+r-i\epsilon/2-i\lambda)}{(m+1/2)\epsilon_1+(n+1/2)\epsilon_2-i(a+r-i\epsilon/2-i\lambda)}=\nn\\
=\prod\lm_{m,n\geq 0}\frac{(m+1)\epsilon_1+(n+1)\epsilon_2-i(a+r+i\lambda)}{m\epsilon_1+n\epsilon_2+i(a+r+i\lambda)}
\frac{(m+1)\epsilon_1+(n+1)\epsilon_2+i(a+r-i\lambda)}{m\epsilon_1+n\epsilon_2-i(a+r-i\lambda)}\sim\nn\\
\mathop{\sim}_{m,n=0}\frac{1}{(a+r)^2-\lambda^2}\mathop{\sim}_{\lambda\rightarrow 0}\delta(a+r)
\ee
one finally obtains
\be
{\cal M}(a,a'|0)\rightarrow \frac{\sqrt{2}\cos\left(4\pi i\frac{aa'}{\epsilon_1\epsilon_2}\right)}{\mu'(a)}
\ee

\subsection*{A3. Chern-Simons average}

Standard quantities arising in the Chern-Simons theory often use
another definition of the dilogarithm. For instance,
the $4_1$-knot average is usually written as a function of the coupling constant $h$ and the knot
monodromy parameter $u$ as (note that $\hbar$ in our formulae differs from $h$ in
\cite{HI} by a factor of 2)
\be
\langle 4_1\rangle=H(u,\hbar)=\frac{1}{\sqrt{\pi \hbar}}\int dp
\frac{\Phi_\hbar(p+i\pi +\hbar/2)}{\Phi_\hbar(-2u-p-i\pi-\hbar/2)}e^{-\frac{4}{\hbar}u(u+p)-u}
\ee
where
\be
\Phi_\hbar(z)=\Phi\left(\frac{z}{\pi i \hbar}\Big| \frac{\hbar}{2\pi i }\right),\quad
\Phi(z|\tau)=\exp\left(\frac{1}{4}\int \frac{d w}{w}\frac{e^{2xz}}{\sinh w\sinh \tau w} \right)
\ee
In the previously discussed context of CFT one can encounter a similar function though with
rescaled parameters
\be
e_b(z)=\exp\left(\frac{1}{4}\int\frac{d w}{w}\frac{e^{-2i zw}}{\sinh b w\sinh b^{-1}w}\right)=\Phi(-ibz|b^2)
\ee
so that
\be
\langle 4_1\rangle=\frac{1}{\sqrt{2\pi (\pi i b^2)}}\int d p
\frac{e_b\left(\frac{1}{2\pi b}\left(p+i\pi +\pi i b^2\right)\right)}
{e_b\left(\frac{1}{2\pi b}\left(-2u-p-i\pi -\pi i b^2\right)\right)}e^{-\frac{2}
{\pi i b^2}u(u+p)-u}
\ee
Introducing new variables $p=2\pi b z$, $u=2\pi b u'$, $Q=b+b^{-1}$, one obtains
\be
\langle 4_1\rangle=\frac{2\pi b}{\sqrt{2\pi(\pi i b^2)}}\int dz
\frac{e_b(z+iQ/2)}{e_b(-2u'-z-i Q/2)}e^{8\pi i u'(u'+z)-2\pi b u'}
\ee
Notice that, though similar, the functions $e_b$ and $S_b$ are slightly different,
in particular, they differ by a multiplier:
\be
e_b(z)=e^{\frac{\pi i z^2}{2}}e^{-\frac{i\pi (2-Q^2)}{24}}S_b(z)
\ee
Therefore,
\be
\langle 4_1\rangle= \sqrt{-2i}\int dz \,\frac{S_b(z+iQ/2)}{S_b(-2u'-z-i Q/2)}e^{6\pi i u'(z+u')-\pi u'(b-b^{-1})}=\nn\\
\mathop{=}_{z\rightarrow z-u'-iQ/2}\sqrt{-2i}\int dz\, S_b(z-u')S_b(z+u')e^{6\pi i u' z+2\pi u'/b+4\pi u' b}
\ee
At last, the same expression in terms of $s(z|\epsilon_1,\epsilon_2)$-dilogarithms is
\be
H\left(\frac{2\pi u}{\sqrt{\epsilon_1\epsilon_2}},\pi i \frac{\epsilon_2}{\epsilon_1}\right)=
\sqrt{\frac{-2i}{\epsilon_1\epsilon_2}}\int dz \, s(z-u|\epsilon_1,\epsilon_2)
s(z+u|\epsilon_1,\epsilon_2)e^{\frac{6\pi i uz }{\epsilon_1\epsilon_2}+\frac{2\pi u}{\epsilon_2}+\frac{4\pi u }{\epsilon_1}}
\ee

\subsection*{A4. Pochhammer symbols as dilogarithm ratios}

Consider the Pochhammer symbols
\be
(q,N)_k=\prod\lm_{j=1}^k\left(q^{\frac{N-j}{2}}-q^{-\frac{N-j}{2}}\right)=
2^k\prod\lm_{j=1}^k \sinh\left(\pi i \hbar (N-j)\right),\nn\\
(q,N)^*_k=\prod\lm_{j=1}^k\left(q^{\frac{N+j}{2}}-q^{-\frac{N+j}{2}}\right)=
2^k\prod\lm_{j=1}^k \sinh\left(\pi i \hbar (N+j)\right)
\ee
Using the periodicity conditions (\ref{s_rel}) one can rewrite the sine products through
\be
\sinh\left(\frac{\pi i \epsilon_2 z}{\epsilon_1}\right)=-\frac{i}{2}\frac{s\left(i\epsilon_2 z +\frac{i}{2}(\epsilon_1-\epsilon_2)\right)}{s\left(i\epsilon_2 z +\frac{i}{2}(\epsilon_1+\epsilon_2)\right)}
\ee
i.e. the Pochhammer symbols are
\be
\label{psim}
(q,N)_k=(-i)^k\prod\lm_{j=1}^k \frac{s\left(i\epsilon_2(N-1-j)+i\epsilon/2\right)}{s\left(i\epsilon_2(N-j)+i\epsilon/2\right)}=
(-i)^k\frac{s\left(i\epsilon_2(N-1-k)+i\epsilon/2\right)}{s\left(i\epsilon_2(N-1)+i\epsilon/2\right)}
\ee
and, correspondingly,
\be
(q,N)_k^*=(-i)^k\frac{s\left(i\epsilon_2(N)+i\epsilon/2\right)}{s\left(i\epsilon_2(N+k)+i\epsilon/2\right)}
\ee
Note that $\langle 4_1\rangle$ is expressed through
$(q,N/\sqrt{\hbar})_{k/\sqrt{\hbar}}$, which
celebrates the symmetry $\hbar\to\hbar^{-1}$:
\be
\langle 4_1\rangle =\frac{i}{\cosh\left(i\epsilon_2(N-1/2)+i\epsilon/2\right)}
\sum\lm_{k\in{\mathbb Z}+\frac{1}{2}}(-1)^k\frac{s\left(i\epsilon_2(N-1/2-k)+
i\epsilon/2\right)}{s\left(i\epsilon_2(N-1/2+k)+i\epsilon/2\right)}
\ee
In the limit $\epsilon_2\rightarrow 0$, $N\rightarrow\infty$,
$i\epsilon_2(N-1/2)+i\epsilon/2=\tilde u$ the sum can be substituted by the integral to give
\be
\langle 4_1\rangle =-\frac{1}{\epsilon_2 \cosh \tilde u}\int d z e^{\pi z/\epsilon_2}s(z-\tilde u) s(z+\tilde u)
\ee
which is quite similar to the Hikami formula, though the correct exponential factor is not
restored in the integrand.

\section*{Appendix B. Examples of knots invariants}

In this section we provide a few simplest examples of knots
and all the associated elements of their description,
outlined in section 3:
for each knot $K$ we list the corresponding
annihilating operator polynomial ${\cal A}$, the spectral curve,
the braid-group element responsible for the quantum-$R$-matrix
description of $<K>_R$, the relevant combination of classical
$R$-matrix and Drinfeld associator, the expansion of $F = \log<K>_R$
in the Vassiliev invariants, the representation of $<K>_R$ and the Hikami
integral $H(K|u)$ through relevant combinations of
$s$-functions.

\subsection*{B1. Unknot}

\begin{itemize}
\item \textbf{Quantum R-matrix representation}

The simplest braid representation of the unknot $U_{0}$ is the closure of single string,
i.e. the closure of the only element of group $B_{1}$:
\be
b_{U_{0}}=1\in B_{1}
\ee
Thus, the value of polynomial invariant is given by the following character:
$$
<U_{0}>=\qtr_{R}(1)=\tr_{R}(q^{\rho})=\chi_{R}\Big(z_{i}=q^{N-2i+1}\Big)
$$
where $\chi_{R}$ is the character of representation $R$ (the corresponding Schur  polynomial).
\item \textbf{Representation through Drinfeld associator}

Formula (\ref{KI}) for the unknot gives vev of unknot in the form of "hump" trace of
the first Drinfeld associator $\Phi_{3}$. Relation between the two is non-trivial: the
$<\hbox{hump}>$
is rather inverse of $<U_0>$. The associator is a tensor with six indices,
and the hump trace is defined as the following contraction:
\be
\label{ch}
<\hbox{hump}>=\sum\limits_{i,k,m=1}^{\dim_{R}}\Big(\Phi_{3}\Big)^{i,i,k}_{k,m,m}
\ee
In the case of fundamental representation of $SU(N)$ this sum was computed explicitly,
and the result can be represented as a particular value of the hypergeometric function
(see \cite{DBetal} for some details):
$$
<\hbox{hump}>=\dfrac{N}{F\Big([(N-1)\hbar, (N+1) \hbar],[ 1+N\hbar], 1\Big)}=
{N(q^{1\over 2}-q^{-{1\over 2}})\over q^{N\over 2}-q^{-{N\over 2}}}={N\over [N]_q}
$$
\item \textbf{Representation through Vassiliev invariant}

The first Vassiliev invariants of unknot are:
\be
\label{uv}
\begin{array}{|c|c|c|c|c|c|c|}
\hline
\alpha_{2,1} & \alpha_{3,1}& \alpha_{4,1} &\alpha_{4,2}&\alpha_{5,1} &\alpha_{5,2}&\alpha_{5,3}\\
\hline
&&&&&&\\-\frac{2}{3}&0&\frac{2}{45}&-\frac{2}{45}&0&0&0\\
&&&&&&
\\
\hline
\end{array}
\ee
For example the usage of (\ref{cfun}) and table (\ref{uv}) for the group $SU(2)$ with spin $J$
gives\footnote{Hereafter, we use the expansions in parameter $h:=2\pi i \hbar$ such that $q=e^{h}$.}:
\be
<U_{0}>_{J}=N \exp\left( h^2 \dfrac{2}{3} J( J+1)- h^{4} {\frac {2}{45}}\, \left( J+1 \right)  \left( 2\,{J}^{2}+2\,J+1
 \right) J +...\right)=\dfrac{q^{{N\over 2}}-q^{-{N\over 2}}}{q^{1\over 2}-q^{-{1\over 2}}}=[N]_q
\ee
with $N=2J+1$.
\item \textbf{${\cal{A}}$-polynomial}

The colored Jones polynomial of unknot
\be
K_N(U_0|q)=<U_{0}>_{J}=\dfrac{q^{J+{1\over 2}}-q^{-(J+{1\over 2})}}{q^{1\over 2}-q^{-{1\over 2}}}
=\dfrac{q^{{N\over 2}}-q^{-{N\over 2}}}{q^{1\over 2}-q^{-{1\over 2}}}
\ee
is a character (\ref{ch}), and, therefore, it satisfies the quantum Laplace equation on the
Cartan lattice:
\be
K_{N+1}-[2]_{q}\, K_N+ K_{N-1}=0
\ee
Therefore, the ${\cal{A}}$-polynomial for the unknot can be defined as
\be
{\cal{A}}(l,m)=(l-1)^2/2,
\ee
and the quantum ${\cal{A}}$-polynomial is the $q$-Laplace operator
\be
\hat{\cal{A}}=\Delta_{q}=\hat l-[2]_{q}+\hat l^{-1}, \ \ \Delta_{q} K_N=0
\ee
where operators $\hat l$ and $\hat m$ act on the Jones polynomial as
\be
\hat l K_N=K_{N+1},\ \ \ \ \ \ \ \ \hat m K_N=q^{N}K_N
\ee

\newpage

\item \textbf{Polynomial invariants}

\paragraph{Colored HOMFLY polynomial:}
\be
\left<U_0\right>_{Y}=\dim_q(Y)=\prod_{i,j\in Y}{q^{{N+j-i\over 2}}-q^{-{N+j-i\over 2}}\over
q^{h(i,j)\over 2}-q^{-{h(i,j)\over 2}}}
\ee
see \cite{OV}. Here $Y$ is the Young diagram corresponding to the representation of $SU(N)$
and $h(i,j)$ is the hook length of a box in $Y$.

\paragraph{Superpolynomial:}
\be
P_0(a,q,t)={a^{1\over 2}-a^{-{1\over 2}}\over q^{1\over 2}-q^{-{1\over 2}}}
\ee

\paragraph{Colored superpolynomial:}
\cite[formulae (67)-(68)]{GIKV}.

\end{itemize}

\subsection*{B2. Knot $3_1$ }

\begin{itemize}
\item \textbf{Representation through quantum $R$-matrix and Drinfeld associator}

The polynomial invariant for $3_1$ can be constructed using formulae (\ref{qi}) and (\ref{KI}).
In this case, $3_1$ is the closure of the following element:
\be
b_{3_1}=g_{1}^3\in B_{2}
\ee
In this representation the knot has $3$ positively oriented crossings ($w(b)=3$) and two strings
($n=2$ in (\ref{KI})).

\item \textbf{Representation through Vassiliev invariant}

The first Vassiliev invariants of knot $3_1$ are:
\be
\begin{array}{|c|c|c|c|c|c|c|}
\hline
\alpha_{2,1} & \alpha_{3,1}& \alpha_{4,1} &\alpha_{4,2}&\alpha_{5,1} &\alpha_{5,2}&\alpha_{5,3}\\
\hline
&&&&&&\\4&-8&\frac{62}{3}&\frac{10}{3}&-\frac{176}{3}&-\frac{32}{3}&-8\\
&&&&&&
\\
\hline
\end{array}
\ee
For example the usage of (\ref{cfun}) for $SU(2)$ gives:
\be
<3_1>=[2J+1]_q\exp\Big( -4 J (1 + J) h^2+8 J (1 + J) h^3+ \frac{2}{3} J (1 + J) (-31 + 10 J + 10 J^2) h^{4}+...  \Big)
\ee
with $h=2\pi i \hbar$.
\item \textbf{Representation through quantum dilogarithm}

The colored Jones polynomial for $3_1$ with $N=2J+1$ can be represented in the hypergeometric form:
\be\label{J3_1}
K_N(3_1|q)=[N]_q \sum\limits_{i=0}^{N-1} (-1)^i q^{\frac{i(i+3)}{2}} (q,N)_{i} (q,N)_{i}^{\ast}
\equiv [N]_qJ_N(3_1|q)
\ee
The (\ref{psim}) gives:
\be
J_N\sim \sum\limits_{k=0}^{N-1}  q^{\frac{k(k+3)}{2}}\, \dfrac{s\left( i \epsilon_2 (N-1-k) + \frac{i}{2} (\epsilon_1+\epsilon_2) \Big| \epsilon_1, \epsilon_2 \right)}{s\left(i \epsilon_2 (N-1) + \frac{i}{2} (\epsilon_1+\epsilon_2) \Big| \epsilon_1, \epsilon_2 \right)}\,\dfrac{s\left( i \epsilon_2 N + \frac{i}{2} (\epsilon_{1}+\epsilon_{2}) \Big| \epsilon_1, \epsilon_2 \right)}{s\left( i \epsilon_2 (N+k) +\frac{i}{2} (\epsilon_{1}+\epsilon_{2}) \Big| \epsilon_1, \epsilon_2 \right)}
\ee
In the limit of $N\to\infty$ and $|q|>1$ this expression turns into
\be\label{3_1Ninf}
J_N\sim q^{{3\over 2}N^2-{3\over 4}N}
\ee

\item \textbf{${\cal{A}}$-polynomial}

The ${\cal{A}}$-polynomial and the spectral curve for the knot $3_1$ have the following form:
\be
{\cal{A}}_{3_1}(l,m)=m^3+l,\ \ \ \Sigma_{3_1}: \{ (m,l)\in {\mathbb{C}}^2: m^3+l=0\}
\ee
Note that this spectral curve corresponds to the sphere, and to the open relativistic Toda
system.

In order to compute the quantum ${\cal{A}}$-polynomial, let us note that the Jones polynomial
for the trefoil (\ref{J3_1}) satisfies the difference equation
\be
J_{N+1}+q^{3N+2}{1-q^N\over 1-q^{N+1}}J_N=q^{N}{q^{2N+1}-1\over q^{N+1}-1}
\ee
which can be rewritten as
\be\label{111}
{1\over \hat B(\hat m)}\hat A(\hat l, m) J_N(q)=1
\ee
with
\be
\hat A =q\hat m^3(\hat m-1)+(q\hat m-1)\hat l,\ \ \ \ \ \ \ \hat B=(q\hat m^2-1)\hat m
\ee
Equivalently, (\ref{111}) can be rewritten as
\be\label{22}
\sqrt{{q\over \hat m^3}}{1-\hat m\over 1-q\hat m^2}
\left(\hat l+q^{3\over 2}\hat m^3\right)K_N(q)=1
\ee
In the leading order at the large $N$ limit this equation reduces to
\be\label{qA}
\left(\hat l+q^{3\over 2}\hat m^3\right)K_N(q)=0
\ee
i.e. the quantum ${\cal{A}}$-polynomial is\footnote{In the literature, one often makes the
different choice of variables: $q\to q^2$ and $\hat m\to \hat m^2$.}
\be
\hat {\cal{A}}_{3_1}(\hat l ,\hat m|q)= \hat l+q^{3\over 2} \hat m^3
\ee
(\ref{3_1Ninf}) definitely solves (\ref{qA}).

Note that at finite $N$ (\ref{22}) can be rewritten as
\be
\hat {\cal A}_gJ_N(q)\equiv(\hat l -1){1\over \hat B(\hat m)}\hat A(\hat l, \hat m) J_N(q)=0
\ee
\be
\hat {\cal A}_g=-{1\over q\hat m}{1-q^2\hat m\over 1-q^2\hat m^2}\hat l^2-\left[
{1\over \hat m}{1-q\hat m\over 1-q\hat m^2}-q^2\hat m^2{1-q\hat m\over 1-q^3\hat m}\right]\hat l
-q^2\hat m^2{1-\hat m\over 1-q^2\hat m}
\ee

\item \textbf{Polynomial invariants}

\paragraph{Colored HOMFLY polynomial:}
The colored polynomial is known in this case only for $SU(3)$, see
\cite[Theorem 1 of the 1st paper]{AJ3}.

\paragraph{Superpolynomial (non-colored):}
\be
P_{3_1}(a,q,t)=P_0(a,q,t)\left[aq^{-1}+aqt^2+a^2t^3\right]
\ee
Choosing $t=-1$ one obtains from this superpolynomial the HOMFLY polynomial. The specialization
of this latter $a=q^N$ corresponds to $SU(N)$ Chern-Simons theory, while $a=q^2$ gives the
Jones polynomial and $a=q^0=1$ gives the Alexander polynomial.

\end{itemize}

\subsection*{B3. Knot $4_1$}

\begin{itemize}
\item \textbf{Representation through quantum $R$-matrix and Drinfeld associator}

The polynomial invariant for $4_1$ can be constructed using formulae (\ref{qi}) and (\ref{KI}).
In this case, $4_1$ can be represented as the closure of the following element:
\be
b_{4_1}=g_{2}^2 g_{1}^{-1} g_{2} g_{1}^{-1}\in B_{3}
\ee
In this representation the knot has $3$ positively oriented crossings and two negative ones
($w(b)=1$) and two strings ($n=3$ in (\ref{KI})).

\item \textbf{Representation through Vassiliev invariant}

The first Vassiliev invariants of knot $4_1$ are:
\be
\begin{array}{|c|c|c|c|c|c|c|}
\hline
\alpha_{2,1} & \alpha_{3,1}& \alpha_{4,1} &\alpha_{4,2}&\alpha_{5,1} &\alpha_{5,2}&\alpha_{5,3}\\
\hline
&&&&&&\\-4&0&\frac{34}{3}&\frac{14}{3}&0&0&0\\
&&&&&&
\\
\hline
\end{array}
\ee
For instance, the usage of (\ref{cfun}) for $SU(2)$ gives:
$$
<4_1>=[N]_q\exp\left\{ J (J + 1) h^2\left[4 + {2\over 3}(14 J^2 + 14 J-17) h^2+
{1\over 90}(2416 J^4+4832 J^3-9212J^2-11628 J+8013)h^4+\right.\right.
$$
\be
\left.\left.+
{1\over 1260}(109552 J^6+328656 J^5-973888 J^4-2495536 J^3+1783146 J^2+3085690 J-1645097)
h^6+
\ldots \right]\right\}
\ee
Naively, in the limit, where $u=hN$ is finite, while $N\rightarrow \infty$
(as usual, $N=2J+1$) only the terms of degree $0$ in $N$ survive, and the series gets
the form
\be
\lim\limits_{N\rightarrow \infty} \ln\Big( <4_1> \Big)= 1+u^2+\frac{7}{12} u^4+ \frac{151}{360} u^6+..= \sum\limits_{n=0}^{\infty} \dfrac{2 u^{2n}}{(2 n)!} \sum\limits_{k=0}^{\infty} \dfrac{k^{2n-1}} {\left(\frac{1+\sqrt{5}}{2}\right)^2}=-\ln\Big( (1-\frac{m}{\gamma})(1-\frac{1}{\gamma m})  \Big)
\ee
where $\gamma=(3+\sqrt{5})/2$ and $m=e^{u}$. No $1/\hbar$-contribution that would correspond
to genus $1/2$ in the t'Hooft expansion is seen in this way.

\item \textbf{Representation through quantum dilogarithm}

The colored Jones polynomial for $4_1$ with $N=2J+1$ can be represented in the hypergeometric form:
\be
K_N(4_1|q)=[N]_q\sum\limits_{i=0}^{N-1}  (q,N)_{i} (q,N)_{i}^{\ast}\equiv [N]_qJ_N(4_1|q)
\ee
Then, (\ref{psim}) gives:
\be
\label{pch4}
K_N \sim \sum\limits_{k=0}^{N-1}  (-1)^k\, \dfrac{s\left( i \epsilon_2 (N-1-k) + \frac{i}{2}
(\epsilon_1+\epsilon_2) \Big| \epsilon_1, \epsilon_2 \right)}{s\left(i \epsilon_2 (N-1) +
\frac{i}{2} (\epsilon_1+\epsilon_2) \Big| \epsilon_1, \epsilon_2 \right)}\,\dfrac{s\left( i
\epsilon_2 N  +\frac{i}{2} (\epsilon_{1}+\epsilon_{2}) \Big| \epsilon_1, \epsilon_2
\right)}{s\left( i \epsilon_2 (N+k) +\frac{i}{2} (\epsilon_{1}+\epsilon_{2}) \Big| \epsilon_1,
\epsilon_2 \right)}
\ee

\item \textbf{Hikami Representation}

To proceed from the previous formula (\ref{pch4}) to the Hikami integral representation,
one needs to go to the double limit $\epsilon_2\rightarrow 0$, $N\rightarrow\infty$,
$i\epsilon_2(N-1/2)+i\epsilon/2=\tilde u$. In this limit, the sum in (\ref{pch4}) should be
replaced by the integral, and one finally gets
\be
\label{tdl}
K_N =-\frac{1}{\epsilon_2 \cosh \tilde u}\int d z e^{\pi z/\epsilon_2}s\Big(z-\tilde u\Big| \epsilon_1, \epsilon_2\Big) s\Big(z+\tilde u\Big| \epsilon_1, \epsilon_2\Big)
\ee
In Hikami's terms, the presence of two dilogarithm functions in the integral (\ref{tdl}) shows
that the hyperbolic space $S^{3}\setminus4_1$ can be obtained by gluing two tetrahedrons.
This representation of $S^{4}\setminus 4_1$ is well studied in the literature, see, e.g.,
\cite{Fuji}.
\item \textbf{${\cal{A}}$-polynomial}

The ${\cal{A}}$-polynomial and the spectral curve for the knot $4_1$ have the following form:
\be
{\cal{A}}_{4_1}(l,m)=m^2 +l(-1+m+2m^2+m^3-m^4)+l^2 m^2,\ \ \ \Sigma_{4_1}=\{ (m,l)\in {\mathbb{C}}^2: {\cal{A}}_{4_1}(l,m)=0 \}
\ee
In order to calculate the quantum ${\cal{A}}$-polynomial in this case, we find the
difference equation (\ref{111}) for knot $4_1$ with
$$
\hat A=q^2\hat m^2(1-\hat m)(1-q^3\hat m^2)-(q\hat m+1)(1-q\hat m-q\hat m^2-q^3\hat m^2-
q^3\hat m^3+q^4\hat m^4)(1-q\hat m)^2\hat l+
$$
\be
+q^2\hat m^2(1-q\hat m^2)(1-q^2\hat m)\hat l^2,\ \ \ \ \ \ \ \ \hat B=q\hat m(1-q^3
\hat m^2)(1-q\hat m^2)(1+q\hat m)
\ee
At the large $N$ limit it reduces to
\be
\left[q^{3/2}\hat m^2(1-q^3\hat m^2)-(1-q^2\hat m^2)(1-q\hat m-q\hat m^2-q^3\hat m^2-q^3\hat m^3+
q^4\hat m^4)\hat l+
\right.\nn\\+\left.
q^{5/2}\hat m^2(1-q\hat m^2)\hat l^2\right]K_N(q)
=0
\ee
and is solved by formula (\ref{tdl}). Thus, the quantum ${\cal A}$-polynomial is
\be
\nn
\hat {\cal{A}}_{4_1}(\hat l ,\hat m|q)=q^{3/2}\hat m^2(1-q^3\hat m^2)-(1-q^2\hat m^2)(1-q\hat m-q\hat m^2-q^3\hat m^2-q^3\hat m^3+
q^4\hat m^4)\hat l+
q^{5/2}\hat m^2(1-q\hat m^2)
\ee
At finite $N$ the Jones polynomial satisfies the equation
\be
\hat {\cal A}_gJ_N(q)\equiv(\hat l -1){1\over \hat B(\hat m)}\hat A(\hat l, \hat m) J_N(q)=0
\ee
$$
\hat {\cal A}_g=-{q^2\hat m(1-q^3\hat m)\over (1+q^2\hat m)(1-q^5\hat m^2)}\ \hat l^3 +
$$
\be
+{1\over q^2}{(1-q^2\hat m)(1+q\hat m-2q^2\hat m-q^3\hat m^2+q^4\hat m^2-q^5\hat m^2-2q^6\hat m^3+q^7\hat m^3+q^8\hat m^4)
\over \hat m(1+q\hat m)(1-q^5\hat m^2)}\ \hat l^2-
\ee
$$
-
{1\over q}{(1-q\hat m)(1-2q\hat m+q^2\hat m-q\hat m^2+q^2\hat m^2-q^3\hat m^2+q^2\hat m^3-2q^3
\hat m^3+q^4\hat m^4)\over
\hat m(1+q^2\hat m)(1-q\hat m^2)}\ \hat l+{q\hat m(1-\hat m)\over (1+q\hat m)(1-q\hat m^2)}
$$

\item \textbf{Polynomial invariants}

\paragraph{Superpolynomial:}
\be
P_{4_1}(a,q,t)=P_0(a,q,t)\left[a^{-1}t^{-2}+q^{-1}t^{-1}+1+qt+at^2\right]
\ee

\end{itemize}

\subsection*{B4. Knot $5_2$}

\begin{itemize}
\item \textbf{Representation through quantum $R$-matrix and Drinfeld associator}

The polynomial invariant for $5_2$ can be constructed using formulae (\ref{qi}) and (\ref{KI}).
In this case, the knot can be represented as the closure of the following element:
\be
b_{5_2}=g_{2}^3 g_{1} g_{2}^{-1} g_{1}\in B_{3}
\ee
In this representation, the knot has $5$ positively oriented crossings and one negatively
oriented  ($w(b)=4$) and two strings ($n=3$ in (\ref{KI})).

\item \textbf{Representation through Vassiliev invariant}

The first Vassiliev invariants of knot $5_2$ are:
\be
\begin{array}{|c|c|c|c|c|c|c|}
\hline
\alpha_{2,1} & \alpha_{3,1}& \alpha_{4,1} &\alpha_{4,2}&\alpha_{5,1} &\alpha_{5,2}&\alpha_{5,3}\\
\hline
&&&&&&\\8&-24&\frac{268}{3}&\frac{44}{3}&-368&-64&-56\\
&&&&&&
\\
\hline
\end{array}
\ee
For instance, the usage of (\ref{cfun}) for $SU(2)$ gives:
\be
<5_2>=[N]_q\exp\Big(  -8 J (1 + J) h^{2}+24 J (1 + J) h^{3}+ \frac{4}{3} J (1 + J) (-67 + 22 J + 22 J^2) \hbar^{4}+... \Big)
\ee
with $h=2\pi i \hbar$.
\item \textbf{Representation through quantum dilogarithm}

The colored Jones polynomial for $5_2$ with $N=2J+1$ can be represented in the hypergeometric form:
\be
K_N(5_2|q)=[N]_q\sum\limits_{0\leq<k \leq l\leq N-1} \dfrac{(q,N)_{l}
(q,N)_{l}}{(q,N)_{k}^{\ast}} \,q^{-k(l+1)}\equiv [N]_qJ_N(5_2|q)
\ee
Then, (\ref{psim}) gives:
$$
J_N \sim \sum\limits_{0\leq<k \leq l\leq N-1} (-i)^{k+2 l} q^{-k(l+1)} \dfrac{s\left( i \epsilon_2 (N-1-l) + \frac{i}{2} (\epsilon_1+\epsilon_2) \Big| \epsilon_1, \epsilon_2 \right)^2 s\left( i \epsilon_2 (N+k) +\frac{i}{2} (\epsilon_{1}+\epsilon_{2}) \Big| \epsilon_1, \epsilon_2 \right)}{s\left(i \epsilon_2 (N-1) + \frac{i}{2} (\epsilon_1+\epsilon_2) \Big| \epsilon_1, \epsilon_2 \right)^2 s\left( i \epsilon_2 N + \frac{i}{2} (\epsilon_{1}+\epsilon_{2}) \Big| \epsilon_1, \epsilon_2 \right)}
$$

\be
\label{hik}
J_N=\dfrac{i}{s(u)^3} \int dz_{1}\int dz_{2} s\Big(u-z\Big| \epsilon_1, \epsilon_2\Big)^2 s\Big(u+z\Big| \epsilon_1, \epsilon_2\Big)\, e^{ -\frac{ 2 \pi z_{1} z_{2}}{\epsilon_{1} \epsilon_{2}} +\frac{\pi i (z_1-z_2)}{\epsilon_1}+\frac{2 \pi i (2z_1+z_{2})}{\epsilon_2} }
\ee
Again, the presence of three dilogarithms in the integral (\ref{hik}) implies that the space
$S^{3}\setminus5_2$ can be realized by gluing three tetrahedrons in the Hikami state model.

\item \textbf{${\cal{A}}$-polynomial}

In this case, one has
\be
{\cal{A}}(l,m)=1+l(-1+2m+2m^2-m^4+m^{5})+l^2(m^2-m^3+2m^{5}+2m^{6}-m^{7})+l^3m^{7}
\ee

The quantum ${\cal{A}}$-polynomial can be calculated as in the previous cases, the result reads
\be
\hat{\cal{A}}(l,m)=q^{1\over 2} (1 - q^4 \hat m^2 )(1- q^{5} \hat m^2) \nn
- (1 - q^2 \hat m^2)(1- q^{5} \hat m^{2})(1 - 2 q  \hat m - q(q + q^3) \hat m^2+\\ \nn
+ q^2(1 - q)(1 - q^2) \hat m^3 + q^{5} \hat m^4 - q^{6} \hat m^{5})\, \hat l\,+
q^{5\over 2} (1 - q \hat m^2)(1 - q^4 \hat m^2) \hat m^2 (1 - q^2 \hat m - q^2(1 - q)(1 - q^2) \hat m^2+\\ \nn
+ q^4(1 + q^3) \hat m^3 + 2q^{7} \hat m^4 - q^{9} \hat m^{5})\, \hat l^2+
q^{14}(1- q \hat m^2)(1-q^{2} \hat m^{2}) \hat m^{7}\, \hat l^3
\ee
This is the simplest example, when $\hat{\cal{A}}(l,m)$ is not quadratic, but cubic in $l$.
In this case the spectral curve is not hyperelliptic.

\item \textbf{Polynomial invariants}

\paragraph{Superpolynomial:}
\be
P_{5_2}(a,q,t)=P_0(a,q,t)\left[aq^{-1}+at+aqt^2+a^2q^{-1}t^2+a^2t^3a^2qt^4+a^3t^5\right]
\ee
\end{itemize}

\section*{Appendix C. Examples of the volume conjecture}

The volume conjecture states that, at large $N$, logarithm of the colored Jones polynomial of
the hyperbolic knot $K$ is proportional to the volume of the knot complement $S^3 \setminus K$:
\be
\log |J_{N}(K)| \sim \frac{N}{2 \pi } Vol \Big( S^{3}\setminus K \Big) \ \ \ \textrm{for} \ \ \ q=e^{2\pi i/N},\ \ \textrm{and} \ \  J_{N}(K)=<K>/\dim_{q}(R)
\ee
Here we represent the plots of $\log |J_{N}(K)|$ for $N=1..100$, for the hyperbolic knot $4_1$
and for the toric knot $3_1$.

For the knot $4_1$, formula (\ref{tdl}) for $q=e^{2\pi i/N}$ gives
\be
J_{N}(4_1)=\sum\limits_{k=1}^{N-1}\prod\limits_{j=1}^{k} 4 \sin^{2} \frac{\pi j}{N}
\ee
Note that all the terms in this formula are positive. Using representation (\ref{pch4}) and
the expansion of quantum dilogarithm  function (\ref{expan}) at large $N$, this sum might be
approximated by the integral
\be
J_{N}(4_1)\sim \int dz \exp\Big( \frac{N}{2\pi i} ( \textrm{Li}_{2}(-e^{-i z})- \textrm{Li}_{2}(-e^{i z}) ) \Big)
\ee
Using the saddle point approximation, one obtains
\be
\frac{d}{dz}( \textrm{Li}_{2}(-e^{-i z})- \textrm{Li}_{2}(-e^{i z}) )=0 \Rightarrow \log( (1+e^{-i z})(1+e^{i z}) )=0 \Rightarrow e^{i z} =e^{\frac{2 \pi i}{3}}
\ee
and for the volume one gets
\be
Vol(4_1)=-i\Big( \textrm{Li}_{2}( -e^{-\frac{2 \pi i}{3}}) -\textrm{Li}_{2}( -e^{\frac{2 \pi i}{3}}) \Big)\approx 2.02688
\ee
One can directly measure this value as a slope in the plot, Fig.3.

In the case of the toric knot $3_1$, the quantity $J_{N}(3_1)$ may take complex values, and,
therefore, the saddle point approximation is more subtle.
Fig.4 shows the power behaviour of the absolute value asymptotics which is
expected \cite{KT} and is consistent with the volume conjecture for the torus knot
$$
|J_{N}(3_1)|\sim N^{3/2}
$$
while the phase of $J_{N}(3_1)$ behaves in a much more tricky way \cite{Mur}.

\begin{figure}
\begin{center}
\includegraphics[scale=0.6, angle=-90]{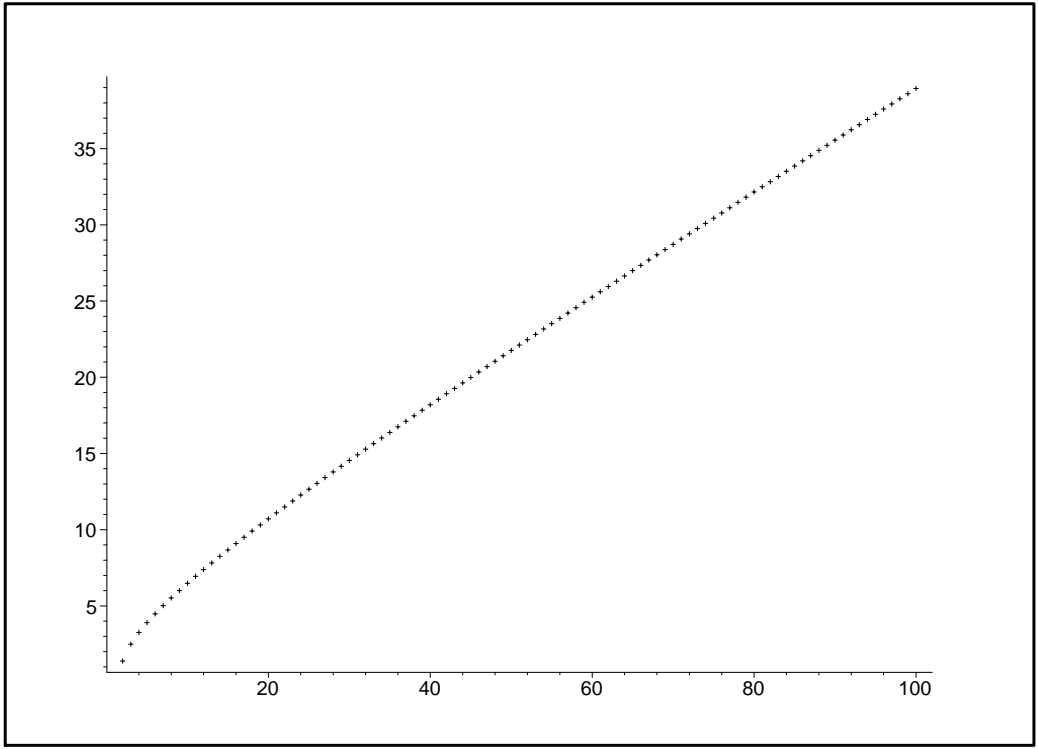}
\caption{The plot of $\ln J_{N}(4_1)$ for $N=3...100$. At large $N$ the plot becomes linear. }
\end{center}
\end{figure}

\begin{figure}
\begin{center}
\includegraphics[scale=0.6, angle=-90]{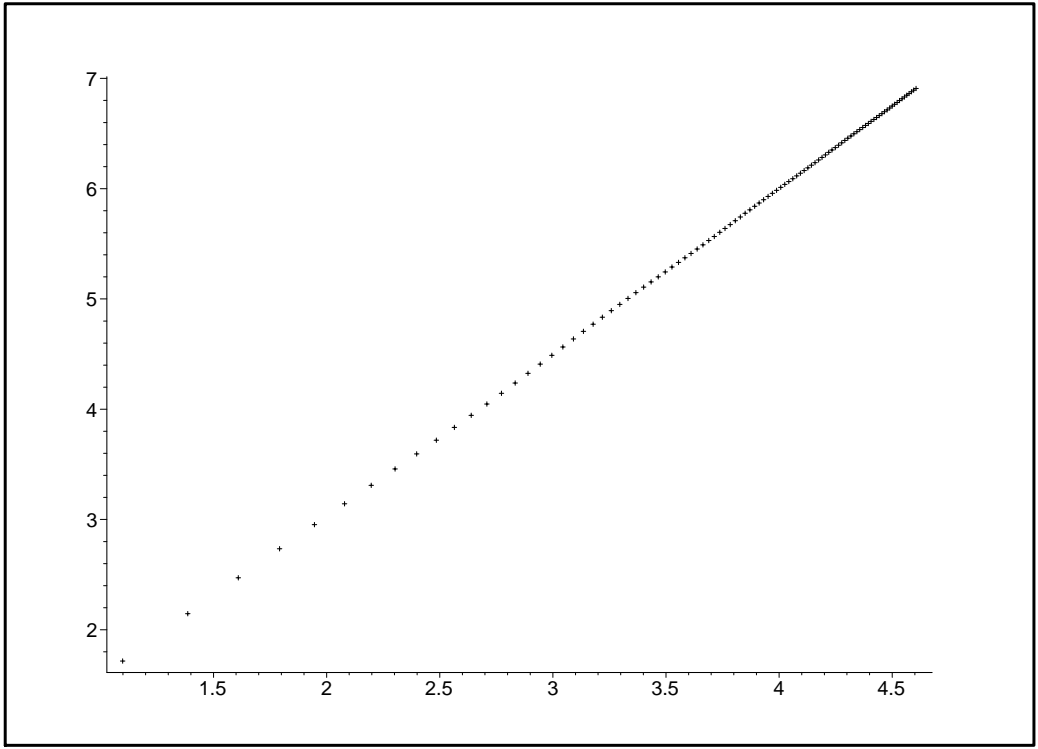}
\caption{The plot of $\ln |J_{N}(3_1)|$ for $N=3...100$ as a function of $\ln N$.
At large $N$ the behavior of the plot is
linear with the slope 3/2.}
\end{center}
\end{figure}

\end{document}